\documentstyle[preprint,aps]{revtex}
\tightenlines
\begin{document}
\draft
\title{A semiclassical approach to the ground state and density 
oscillations of quantum dots}
\author{A. Puente, M. Casas and Ll.\ Serra}
\address{Departament de F\'{\i}sica, Universitat de les Illes Balears,
E-07071 Palma de Mallorca, Spain}
\date{19 August 1999}
\maketitle
\begin{abstract}
A semiclassical Thomas-Fermi method, including a Weizs\"acker gradient
term, is implemented to describe ground states of two dimensional 
nanostructures of arbitrary shape. 
Time dependent density oscillations are addressed 
in the same spirit using the corresponding semiclassical time-dependent
equations. The validity of the approximations is tested, both for ground 
state and density oscillations, comparing with the available microscopic
Kohn-Sham solutions.

\end{abstract}

\narrowtext

\section{Introduction}

Two-dimensional quantum dots, and semiconductor nanostructures in general,
are examples of artificial systems in which, ultimately, we may seek
to produce and control electronic quantum properties. Sometimes they are
referred to as artificial atoms and in fact, it has been proved that the 
electronic structure of circular quantum dots resembles in some aspects the 
shell structure found in atoms and nuclei. Particularly relevant in this 
sense are the Coulomb blockade measurements that provide direct acces to 
the electronic energy levels by adding electrons one by one to the quantum 
dot \cite{Ash96}. The magic numbers measured in vertical quantum dots
\cite{Tar96}, corresponding to maxima in the addition energies, have 
also shown clear evidences of shell structure.
Absorption measurements in the far-infrared region \cite{Sik89,Dem90}
as well as experiments using sophisticated light scattering 
techniques \cite{Loc96,Sch98} have probed 
both charge density (CDE) and spin density (SDE) excitations and 
have shown that parabolic confinement is closely satisfied for small size 
dots.

Interest in the quantum dot community has recently focussed 
on the properties of deformed nanostructures. For instance, elliptic 
quantum dots have been investigated in Refs.\ 
\onlinecite{Aus99,Mad94,Hir99,Yan99,Pue99,Ser99,Mag99}.
Particularly, in Refs.\ \onlinecite{Pue99,Ser99} some of us 
have addressed the collective 
oscillations of deformed dots where, in addition to density 
and spin modes, it has
been predicted the existence of orbital current modes at low energies.
This interest in deformed dots is motivated by the advances in 
nanofabrication techniques, that allow to produce quantum dots of many 
different shapes. 

Microscopic theoretical approaches, like Hartree 
\cite{Bro90}, Hartree-Fock \cite{Gud91} and density functional 
\cite{Fer94,Pi98,Lip98}, let alone exact diagonalizations \cite{Mak90,Yan93}, 
get very demanding computationally for increasing number of electrons,
especially in the symmetry unrestricted case. It is thus interesting to 
develop a semiclassical approach, dealing only with the total density,
for which the computational effort
is not much dependent on the number of electrons. 
This is our purpose in this paper. 
Semiclassical Thomas-Fermi models have been already 
used in the field, 
for instance in Refs.\ \onlinecite{Mce92,Mar92,Lie95,Bar98}. 
In particular, Ref.\ \onlinecite{Lie95}
provides a quite rigorous presentation of the theory. However, 
previous works deal only with the semiclassical ground state and mostly 
for circular symmetry. Here we emphasize the application to deformed 
structures and concentrate on the oscillation modes.
We do not include in this paper magnetic field since this requires a 
non trivial generalization of the theory that we plan to develop in
a separate contribution.

The structure of the paper is as follows. In Sec.\ II we discuss the ground 
state in our semiclassical approach and compare with available microscopic 
Kohn-Sham (KS) results. Section III introduces the formalism for the 
time-dependent oscillations. Results for the oscillation modes are 
discussed in Sec.\ IV. Finally the conclusions are drawn in Sec.\ V.

\section{Ground states}

\subsection{Definition of energy functional}

Using Density Functional Theory in a local approximation we write 
the total energy in terms of the electronic density $\rho({\bf r})$ as
$E[\rho]=\int{d{\bf r}\, {\cal E}[ \rho ]}$, with the energy density 
separating 
as 
\begin{equation}
{\cal E}[ \rho ]=\tau[\rho]+
{1\over 2} v_H({\bf r})\rho +{\cal E}_{XC}(\rho)+
v_{\em ext}({\bf r})\rho\; .
\end{equation}
The different terms are the kinetic energy density $\tau[\rho]$, the Hartree
potential 
$v_H({\bf r})=\int{
d{\bf r}' {\rho({\bf r}'\,)\over|{\bf r}-{\bf r}'\,|}}$, the 
exchange-correlation energy density ${\cal E}_{XC}(\rho)$ and 
the external confining potential $v_{\em ext}({\bf r})$. The Thomas-Fermi
approximation in two dimensions (2D) yields the following kinetic energy density
\cite{Bra98}
\begin{equation}
\tau_{TF}[\rho]= {\hbar^2\over 2m} \left(\pi\rho^2+
{5\over 12}\nabla^2\rho\right)\; .
\end{equation}
By analogy with the Weiz\"acker term for the 3D kinetic energy 
functional, we have added a gradient term 
that gives the exact kinetic energy for a single electron
\begin{equation}
\tau_W[\rho]={\hbar^2\over 2m} \lambda {(\nabla\rho)^2\over\rho}\; ,
\end{equation}
with $\lambda=1/4$. Our total Thomas-Fermi-Weizs\"acker (TFW)
kinetic functional is then 
$\tau=\tau_{TF}+\tau_W$.
It is worth to point out that the first non-vanishing gradient correction
in 2D
is not known from a rigorous semiclassical expansion and therefore 
we have introduced empirically the Weizs\"acker-like term. We will 
show later that the results are not sensitive to the precise value of the 
coefficient $\lambda$. For the sake of comparison, we recall here 
that the KS method provides the exact kinetic energy by means 
of a set of single particle orbitals $\{\varphi_i\}$ as
\begin{equation}
\tau({\bf r})={\hbar^2\over 2m}\sum_{i,{\em occ.}}
{|\nabla\varphi_i({\bf r})|^2} \, .
\end{equation}
For the exchange-correlation energy ${\cal E}_{XC}$ we have used the same 
functional of Ref.\ \onlinecite{Pi98}, based on the Tanatar and Ceperley
calculation for the uniform gas \cite{Tan89}.
Specifically, in modified atomic units \cite{units}, it is written as 
\begin{equation}
{\cal E}_{XC}(\rho)=-{4\over 3}\sqrt{2\over\pi}\,\rho^{3/2} 
+{1\over 2} a_0 \rho\, {1+a_1 x\over 1+a_1 x+a_2 x^2 + a_3 x^3}\; ,
\end{equation}
where $x=(\pi\rho)^{-1/4}$ and the $a_i$ coefficients are given in Ref.\ 
\cite{Tan89}. Notice that in this formalism we always assume perfect 
spin degeneracy in the ground state in order to obtain a single 
chemical potential for both spin components (see below).

The ground state density is determined from the minimization (Euler-Lagrange)
equation, with the contraint of particle number conservation
\begin{equation}
{\delta\over\delta\rho}\left(
E[\rho]-\mu\int{d{\bf r}\, \rho} \right) =0 
\; .
\end{equation}
Specifically, this reads
\begin{eqnarray}
-2\lambda {\hbar^2\over 2m} \nabla^2\rho &+&
\lambda {\hbar^2\over 2m} {(\nabla\rho)^2\over\rho} \nonumber\\
&+& \left(v_{\em ext}+v_H+
{\partial{\cal E}_{XC}\over\partial\rho}+2\pi{\hbar^2\over2m}\rho\right)
\rho=\mu\rho\; .
\end{eqnarray}
By making the transformation $\psi=\sqrt{\rho}$ this equation may be 
written in the familiar form of a Schr\"odinger like equation
\begin{equation}
-4\lambda{\hbar^2\over 2m} \nabla^2\psi + V \psi = \mu \psi\; ,
\label{eq7}
\end{equation}
where we have defined the semiclassical effective potential 
$V=v_{\em ext}+v_H+
{\partial{\cal E}_{XC}\over\partial\rho}+2\pi{\hbar^2\over2m}\rho$.
The chemical potential $\mu$ plays the role of the eigenvalue in 
Eq.\ (\ref{eq7}). Written in this way, we may now use the algorithms 
developed to solve the KS equation in arbitrary 2D confinements
\cite{Pue99}. Our method is based on a discretization of the $xy$ plane 
in a grid of uniformly spaced points. Then, the total number of grid points, 
not the number of electronic orbitals as in KS theory, 
determines the computational cost of the problem.
Typically, we use grids ranging from $50^2$ to $100^2$  points.

\subsection{Results and comparison with Kohn-Sham}

In this subsection we show results for the semiclassical ground states in 
different confining potentials,
focussing especially on the comparison with the corresponding microscopic 
KS results in order to prove the validity of the approximations. 
Since the microscopic calculation for medium and large sizes is 
computationally feasible only when circular symmetry is imposed to the 
system, we will begin by considering the cases of circular parabolas and
disks of jellium. In these cases we compare with the KS radial
solution \cite{Pi98}. Then we analyze one case with deformation, 
namely the deformed parabola with $N=20$, and compare with a symmetry 
unrestricted microscopic calculation \cite{Pue99}.

\subsubsection{Circular parabolas}

The confining potential in this case is 
(we will use modified atomic units \cite{units} for the rest of the paper)
\begin{equation}
v_{\em ext}(r)= A_0 + {1\over 2} \omega_0^2 r^2\; .
\end{equation}
The constants $A_0$ and $\omega_0$ are usually parameterized
to reproduce the potential and its curvature at the origin of a jellium 
disk (see below) with $N_p$ positive charges 
as $A_0 = -2 \sqrt{N_p}/r_s$ and $\omega_0^2 = 1/(\sqrt{N_p}r_s^3)$, 
with $r_s$ the radius per unit charge.
Figure 1 shows the densities obtained taking 
$r_s=1.51 a_0^*$. 
Panels (a) and (b) prove that the semiclassical density closely 
adjusts to the microscopic one for varying electron number 
in a fixed parabola and for a varying parabola curvature 
at fixed electron number, respectively.
As is well known from atomic physics, the TFW density averages the 
microscopic shell oscillations in the inner region. Panel (c) shows that the 
coefficient $\lambda$ controls the surface width of the semiclassical 
density. The value $\lambda=1/4$ provides a reasonable approximation
to the KS density tails in
all cases. However, it is clear that if this coefficient is allowed to 
vary a better fit of the densities may be obtained. We have not followed 
this procedure since the fitted value of $\lambda$ would be different  
for each type of confining potential (see next subsections).

In Fig.\ 2 we compare total energies per electron in the two approaches. 
The fit of energies is amazingly good; the deviation of the TFW 
energies with respect to KS being less than
0.5\% for all cases shown in Fig.\ 2.
Although the high precission 
of the TFW model for this quantity may seem a bit surprising, we 
recall that the method looks for the variational minimum of total 
energy. Therefore, this is a priori the best value of the model.
We have made the comparison for magic electron numbers, corresponding
to closed shell configurations. This is the reason why shell oscillations
with size are not visible in the microscopic results of Fig.\ 2(a).
The total energies are almost independent of $\lambda$ for reasonable 
values of this coefficient. The results for $N=42$ and $\lambda=0.25$, 0.5
and 1.5 are $E/N=-4.033$, -4.025 and  -4.014 H$^*$, respectively. 

\subsubsection{Jellium disks}

Another external potential with circular symmetry is that of a uniformly
charged disk. Defining the positive charge density (jellium density) in
terms of the $r_s$ parameter $\rho_j=1/(\pi r_s^2)$ and for a disk with 
radius $R$, the potential is
\begin{equation}
v_{\em ext}(r)=\left\{ 
\begin{array}{ll}
-{\displaystyle{4\over\pi} 
{R\over r_s^2}} E(r/R) &\qquad {\rm if}\quad r\leq R\\
-{\displaystyle{4\over\pi} {r\over r_s^2}} \left[ 
E(R/r)- \left( 1- \left(\displaystyle{ R\over r}\right)^2 \right) K(R/r)
\right] &\qquad {\rm if}\quad r\geq R
\end{array}
\right.\; ,
\end{equation}
where $E$ and $K$ are the elliptic integrals. Assuming a uniformly 
charged disk the 
number of positive charges $N_p$ is related to the disk radius by
$R=r_s\sqrt{N_p}$.

Contrary to the parabola, the jellium potential forces the electronic 
density to saturate inside the dot. This is obviously due to the charge 
screening effect, that energetically favors the cancellation of charges.
Figure 3 shows the electronic densities of neutral disks 
with $r_s=1.51 a_0^*$ as a function of size. 
The agreement between both models is rather good. As for the 
parabolas, the TFW densities average the oscillations of the KS ones 
but now they rapidly 
saturate inside the dot. Close to the edge, the TFW densities present an 
small oscillation. This is similar to the Friedel-type oscillations 
found in metal surfaces and
is enhanced by potentials that abruptly vanish at the edge, as the 
jellium one. Also shown in Fig.\ 3 (panel b) are the energies per electron 
of the neutral disks, that are also very well reproduced by the 
semiclassical model.

From Fig.\ 3 we see that the TFW density tails slightly overestimate the KS 
ones, with the used value of $\lambda$. This implies that the number of 
electrons that are spilling out of the jellium disk will be slightly 
enhanced in the 
semiclassical method. This spill out mechanism is known from cluster 
physics to be of fundamental importance for a proper 
description of the optical properties \cite{Bra93}. 
The fact that our TFW method 
correctly includes spill out gives us some confidence in its use for 
the description of time dependent density oscillation (see Sec.\ IV).

\subsubsection{Deformed parabola for N=20}

The third case we have considered is an elliptical dot, confined by an 
anisotropic parabola, 
\begin{equation}
v_{\em ext}({\bf r}) = {1\over 2} \omega_0^2 {4\over (1+\beta)^2}
(x^2+\beta^2 y^2)\; .
\end{equation}
The parameter $\beta$ gives the ratio of parabola coefficients in $y$ and 
$x$ directions, i.e., writing the external potential as
$v_{\em ext}({\bf r}) = {1\over 2}( \omega_x^2 x^2 + \omega_y^2 y^2 )$,
we have $\beta=\omega_y/\omega_x$. 
At the same time the centroid ${(\omega_x+\omega_y)/2}$ is kept 
fixed at the value $\omega_0$, defined as for the circular parabola 
$\omega_0^2=1/(r_s^3\sqrt{N_p})$.

Since the KS problem is much more involved than for the preceding 
circular potentials, we restrict here the comparison to the $N=20$ electrons
case, with $N_p=20$, and consider four deformations $\beta=0.875$, 0.75, 
0.625 and 0.5. Figure 4 shows the densities corresponding to these four 
deformations as well as the circularly symmetric case ($\beta=1$)
for completeness.
It is seen that, quite nicely, the elliptic contour lines 
of the TFW density follow in average the equidensity regions of the 
KS result. 
For $\beta=1$ and $\beta=0.875$ the structure of equidensity regions 
of both models is very similar. For lower values of $\beta$ the 
microscopic model yields an incipient electron localization that the 
semiclassical model is obviously not able to reproduce; it gives 
nevertheless the correct average value. We conclude that the semiclassical 
model reproduces 
the average density distributions also in non circular systems.
Total energies, given in Tab.\ I, are also in excellent
agreement with the KS ones.

\section{Time dependent equations}

To derive the time dependent equations in the semiclassical approximation
we will follow the fluid dynamical approach. This method has been used
in nuclear physics \cite{Rin80} and, more recently, also for electronic 
oscillations in atomic clusters \cite{Dom98}. The variational derivation of
these equations can be found in the book by Ring and Schuck \cite{Rin80}. 
Here we will just point out the essential ingredients and particular details 
for the application to our case.

The essential assumption is that all the single 
particle orbitals evolve in time with a common complex phase as
\begin{equation}
\varphi_i({\bf r},t)=\varphi_i^{(0)}({\bf r},t)\, e^{is({\bf r},t)}\; ,
\label{eq5.1}
\end{equation}
where both $\varphi_i^{(0)}({\bf r},t)$ and $s({\bf r},t)$ are assumed 
to be real functions. With (\ref{eq5.1}) the kinetic energy 
separates in two contributions 
\begin{equation}
T=T_{\em intr}+{1\over 2}\int{d{\bf r} \rho {\bf u}^2}\; ,
\label{eq5.2}
\end{equation}
an intrinsic one 
$T_{\em intr}={1\over 2}\sum_i{|\nabla\varphi_i^{(0)}|^2}$
and another associated to the common velocity field 
${\bf u}=\nabla s$. Generalizing this separation we may write for 
the total energy (expectation value of the Hamiltonian $H$)
\begin{equation}
{\cal H} \equiv \langle H\rangle =
E_{\em intr} + {1\over 2}\int{d{\bf r} \rho {\bf u}^2}\; .
\label{eq5.3}
\end{equation}
In our density functional approach 
the intrinsic energy is in fact given by the energy functional
$E_{\em intr}=E[\rho]$.
Noticing now that $\rho$ and $s$ are conjugated canonical variables, 
Hamilton's equations are \cite{Rin80}
$\dot{\rho} = {\delta{\cal H}\over\delta s}$, 
$-\dot{s} = {\delta{\cal H}\over\delta\rho}$.
Specifically, these read
\begin{eqnarray}
\dot{\rho} &=& -\nabla(\rho\nabla s)\nonumber\\
-\dot{s} &=& {1\over 2}(\nabla s)^2 + 
{\delta{E[\rho]}\over\delta\rho}\; .
\label{eq5.4}
\end{eqnarray}
The two Eqs.\ (\ref{eq5.4}) yield the time evolution of the semiclassical 
variables $\rho$ and $s$, and thus they are our required input to 
model a semiclassical dynamics in quantum dots. As happens for the ground 
state equation, we may still transform Eqs.\ (\ref{eq5.4}) to a form 
similar to the microscopic equation. In fact, defining
the time dependent complex function $\psi = \sqrt\rho e^{is}$ both Eqs.\ 
(\ref{eq5.4}) reduce to 
\begin{equation}
i{\partial\psi\over\partial t} =
-{1\over 2} \Delta \psi +
\left( {\delta{E[\rho]}\over\delta\rho}+
{1\over 2} {\Delta\sqrt\rho\over\sqrt\rho} \right) \psi\; ,
\label{eq5.5}
\end{equation}
i.e., an equation identical to the time-dependent KS equation 
if we identify the contribution within brackets in the right hand side 
with the potential term. 

In the preceding discussion we have assumed, for simplicity, that both 
spin densities are oscillating in phase, and thus identically to the total 
density. The formalism, however, can also account for the situation in 
which both spin densities $\rho_\eta$ ($\eta=\uparrow, \downarrow$) 
oscillate out of phase, as happens for instance in spin modes. In this case
we just need to assume the semiclassical approximation for each spin fluid,
with the Coulombic coupling terms between them and generalize the 
exchange-correlation contribution to the locally polarized case using the 
exchange interpolation formula \cite{Tan89}.
Defining 
$\psi_\eta = \sqrt\rho_\eta e^{is_\eta}$ the two equations are
\begin{equation}
i{\partial\psi_\eta\over\partial t} =
-{1\over 2} \Delta \psi_\eta +
\left( {\delta{E}[\rho_\uparrow,\rho_\downarrow]\over\delta\rho_\eta}+
{1\over 2} {\Delta\sqrt\rho_\eta\over\sqrt\rho_\eta} \right) \psi_\eta\; .
\label{eq5.6}
\end{equation}

In this paper we apply the time dependent equations to obtain the 
linear response frequencies corresponding to dipole charge density (CDE)
and spin density (SDE) excitations of general 2D nanostructures. These are 
normal modes of oscillation and the technique to obtain them is simply to 
apply an small instant perturbation on the ground state and take this 
as initial 
condition for the time simulation. For the CDE the initial perturbation
is simply a rigid translation of the total electronic density, given 
by operator ${\cal T}({\bf a})$, with ${\bf a}$ the vector displacement.
For the SDE we apply opposite translations for spin up and down densities,
${\cal T}_\eta({\bf a}_\eta)$, with ${\bf a}_\uparrow=-{\bf a}_\downarrow$.
After this, we keep track of the time evolution of the dipole moments
$d_{\eta}(t)=\langle {\bf r} \rangle_\eta\cdot\hat{\bf e}$, where $\hat{\bf e}$ is 
a unitary vector in the direction of the displacement, and
\begin{equation}
\langle {\bf r}\rangle_\eta ={1\over N_\eta} \int{d{\bf r} 
\rho_\eta({\bf r},t) {\bf r}} \; ,
\end{equation}
with $N_\uparrow$, $N_\downarrow$ the number of electrons with 
spin up and down, respectively. A frequency analysis of the total 
dipole moment $d_\uparrow+d_\downarrow$ for the CDE, and of the 
spin dipole moment $d_\uparrow-d_\downarrow$ for the SDE, finally 
yields the oscillation frequencies \cite{Pue99}.

\section{Results of density oscillations}

We present in this section the spectra obtained within the semiclassical 
formalism. We consider density and spin dipole oscillations and, as in 
Sec.\ I, we emphasize the comparison with the corresponding KS results.
One case is taken as representative for the three types of confining 
potentials for which we have already discussed the ground state.

\subsubsection{Circular parabola with $N=56$}

Figure 5 shows the CDE and SDE for the parabola with $N=56$ and $N_p=40$.
The results have been plotted in a logarithmic arbitrary vertical scale 
(each tick marks an order of magnitude).
First, we notice that in the CDE the semiclassical spectrum (and also the 
KS one) yield a single frequency that coincides with the parameter 
$\omega_0$ of the confining potential. We thus conclude that time-dependent
TFW satisfies well the generalized Kohn's theorem \cite{Mak90}. This theorem
states that the exact dipole CDE for parabolic confinement is a single peak
at the parabola frequency $\omega_0$. 
The more intense SDE's lie at lower energy because the residual interaction
is attractive in this channel \cite{Ser97}. The semiclassical model
does not include the coupling to particle-hole excitations, that leads 
to an important fragmentation of the KS spectrum. Nevertheless, it 
reproduces some fragmentation of the collective strength, 
contained in two dominant peaks at 
$\approx 0.05$~H$^*$ and $\approx 0.19$~H$^*$, that nicely 
correspond to very intense KS excitations.

We remark that the 
microscopic results in Fig.\ 5 (and also those of Fig.\ 6 below) have been 
obtained by using the perturbative random-phase approximation (RPA) 
in a particle-hole basis \cite{Lip98} while 
the TFW results were obtained, as explained in Sec.\ IV, from a frequency 
analysis of the real time signal. The RPA calculation provides a very high 
frequency resolution, even for low intensity peaks. On the contrary, 
the analysis of the time signal is not able in some cases 
to discriminate the low intensity peaks because of the limitations of a 
discrete time sampling and a finite total time window.

\subsubsection{Circular jellium with $N=58$}

This is shown in Fig.\ 6. In this case generalized Kohn's theorem does not 
hold since the external potential is not of parabolic type. As a consequence
the dipole charge oscillation couples to the relative motion and the 
spectrum is generally fragmented, with peak energies depending on the number of 
electrons \cite{Ser97}.
The disk potential behaves quadratically close to the disk center but it deviates 
for points closer to the edge. Fig.\ 6 shows the $\omega_0$ value for the 
quadratic behaviour at the origin. The semiclassical model reproduces the blue 
shift from 
$\omega_0$ although it slightly underestimates its quantitative value.
This may be traced back to the slight difference in spill out mentioned in 
Sec.\ II.B.2. The TFW model yields a greater spill out and thus produces a 
softer oscillation mode. 
The absence of an exact Kohn mode manifests with an important fragmentation,
partly reproduced in the TFW model.
In the SDE of Fig.\ 6 the situation is similar to the circular parabola. 
The semiclassical spectrum reproduces the dominant collective peaks, and 
has less fragmentation than KS.

\subsubsection{Deformed parabola}

Figure 7 shows the results for an anisotropic parabola with $N=20$,
$N_p=20$ and different deformations. 
The two upper panels correspond 
to the CDE for $\beta=0.5$ and 0.75. They prove that also in deformed 
parabolas the generalized Kohn's theorem is well satisfied by
time dependent TFW. In this case the parabolas in $x$ and $y$ directions
are different and, in fact, the oscillation in these two directions
is at frequency $\omega_x$ and $\omega_y$, respectively.
For the two other deformations not shown in the figure 
($\beta=0.625$, 0.875) generalized Kohn's theorem is equally satisfied.
Therefore, as for TDLDA, TDTFW fulfills the exact property 
for dipole charge oscillations
saying that the two center of mass coordinates $X=\sum_i x_i$ and
$Y=\sum_i y_i$ oscillate with the frequencies $\omega_x$ and $\omega_y$,
respectively \cite{Dob94}. 

The four lower panels of Fig.\ 7 show the SDE spectra for the four
deformations considered. The $x$-$y$ splitting in the spin channel is 
nicely reproduced, as a function of deformation within the TFW model. 
Fragmentation is present in both models, although TFW is overestimating 
the collective strength at high energy for $\beta=0.75$ and $\beta=0.875$.
We attribute this to the rather small electron number of this dot ($N=20$),
for which TFW is surely less accurate than for large sizes.  We remark 
again that the lower energy (and more intense) peak, as well as the 
$x$-$y$ splitting are correctly reproduced.

To finish this section and in order to emphasize the power of the 
semiclassical method we reproduce in Fig.\ 8 a case that is beyond 
the present capability of microscopic KS calculations. It corresponds
to 72 electrons in a deformed parabola, with $N_p=72$ and $r_s=1.51 a_0^*$. 
Since the CDE corresponds only to $\omega_x$ and $\omega_y$, according to 
Kohn's theorem, we display the result of the SDE for 
$\beta=0.75$, as well as the circular one for comparison.

\section{Conclusions}

In this paper we have discussed a semiclassical approach to the ground
state and density oscillations of 2D nanostructures. 
The method has been 
implemented for the general case in which no spatial symmetry 
is required. The validity of both ground state and dipole oscillation
descriptions has been checked by systematically comparing the 
semiclassical TFW 
results with the corresponding KS ones. We have shown 
that the TFW densities closely follow 
the KS ones, averaging the shell oscillations, for different types
of external confinings: circular and deformed parabolas and jellium disks 
for which the density saturates. Besides, the TFW model reproduces the KS 
energies with great accuracy. 
The dependence of the energy with the value of the 
Weizs\"acker coefficient is very small. This coefficient controls the 
density tail and electronic spill out for jellium disks. We have shown that
the value $\lambda=1/4$ provides a good overall fit. 

Dipole charge density and spin density oscillations
have been analyzed by using the time dependent semiclassical 
equations. 
In circular parabolic 
confinements we have shown that the semiclassical spectra satisfy well 
the generalized Kohn's theorem for charge density excitations, while 
in neutral jellium disks it yields a blue shift
similar to the KS one. In elliptic dots the generalized Kohn's
theorem is satisfied as well.
For spin density excitations the TFW model is able to reproduce the 
dominant peaks of the spectrum and the splitting associated with 
deformation in elliptic dots.

In general, we have shown that the semiclassical Thomas-Fermi-Weizs\"acker 
model provides a reliable tool to obtain accurate approximations to the 
ground state and linear oscillations of quantum dots. This opens the 
possibility to use it in order to explore a great variety of 
confining potentials with different geometries and for large sizes.
\vspace*{1cm}

This work was performed under
Grant No.\ PB95-0492 from CICYT, Spain.

%%Figures
\begin{figure}[h]
\caption{Radial electronic densities for circularly symmetric parabolas in the
TFW model (dashed) in comparison with 
the corresponding KS densities (solid).
Panel (a) corresponds to a fixed parabola 
with $\omega_0=0.21$ H$^*$, that  
with the parameterization used in the text is given by 
$N_p=40$ and $r_s=1.51 a_0^*$, with different number of electrons. Ordered 
vertically at $r=0$ are shown $N=2$, 6, 12, 20, 30, 42, 56, 72 and 90,
respectively.
Panel (b) corresponds to a fixed number 
of electrons $N=42$ and varying parabolas with the same $r_s$ and 
$N_p=20,30,40,50$ and 60 (also ordered vertically at $r=0$) .
Panel (c) shows the density for a fixed dot, $N=40$ and 
$N_p=42$ within KS and for varying Weizs\"acker coefficient 
$\lambda=0.25$, 0.75 and 1.5 (in vertical order around $r=14$).
}
\end{figure}

\begin{figure}[h]
\caption{Energy per electron, as a function of electron number $N$,
in the TFW model (smooth line) and in the KS model (symbols).
Panels (a) and (b) contain the results for the corresponding panels 
in Fig.\ 1.
}
\end{figure}

\begin{figure}[h]
\caption{Results for neutral jellium disks with $r_s=1.51 a_0^*$.
Panel (a) shows the radial densities
within the TFW (dashed) and KS (solid). With increasing edge radius
the results correspond to $N=12$, 20, 46, 58 and 72.
Panel (b) shows the corresponding energies for KS (symbols) and 
TFW (smooth line).
}
\end{figure}

\begin{figure}[h]
\caption{Comparison of KS and TFW densities for anisotropic parabolas
with $N=20$, $N_p =20$ and $r_s =1.51 a_0^*$ for   
different deformation parameters. The elliptic contour lines 
correspond to the semiclassical densities while the coloured background 
regions show the KS results. The scale bar indicates the density associated
with each colour as well as its value for the semiclassical contour lines,
which correspond approximately to the colour transitions.
}
\end{figure} 

\begin{figure}[h]
\caption{Dipole excitation spectra, in logarithmic 
arbitrary units, for charge density 
and spin density excitations in a parabolic quantum dot.
In each panel, the upper curve corresponds to the microscopic 
RPA calculation and the lower one to the present semiclassical 
approximation.
In the upper panel the triangle shows the value of the parabola 
coefficient $\omega_0$.
}
\end{figure}

\begin{figure}[h]
\caption{Same as Fig.\ 4 for a jellium disk with $r_s=1.51a_0^*$. 
In this case the triangle corresponds to the parabola that fits 
the disk potential at the origin.
}
\end{figure}

\begin{figure}[h]
\caption{
Dipole spectra for the parabolic quantum dots of Fig. 4. Solid and dashed lines show 
the results for oscillations in $x$ and $y$ directions, respectively.
For the CDE (two upper panels) we plot the TFW result and the position
of the $\omega_x$ and $\omega_y$ parameters 
while for the SDE (four lower panels) we 
compare with the corresponding KS curves (displaced vertically
above the TFW ones). As before, vertical scales are in 
logarithmic arbitrary units.
}
\end{figure}
\begin{figure}[h]
\caption{
Same as Fig.\ 7 for a parabolic dot with $N=72$ and $N_p=72$.
The microscopic KS results are not available for $\beta=0.75$.
}
\end{figure}

\begin{table}[]
\caption[]{Energies per electron (in H$^*$) within the TFW and KS models. The 
results correspond to a parabola with $N=20$, $N_p=20$, $r_s=1.51 a_0^*$ 
and different deformations. See text.}
\begin{center}
\begin{tabular}{c|ccccc}
 & \multicolumn{5}{c}{$\beta$} \\
       & 1 & 0.875 & 0.75 & 0.625 & 0.5 \\
\hline
KS     & $-3.026$  & $-3.032$ & $-3.059$ & $-3.118$ & $-3.223$  \\
TFW    & $-3.025$  & $-3.033$ & $-3.061$ & $-3.119$ & $-3.225$  \\
\end{tabular}
\end{center}
\end{table}


\begin{references}
\bibitem{units} In modified atomic 
units one imposes: $\hbar=1$, $m=1$ and $e^2/\varepsilon=1$. 
Using $m$ and $\varepsilon$ values corresponding to GaAs the energy, 
length and time units are 
${\rm H}^*\approx 12$~meV, $a_0^*\approx 98$~{\AA} and
$\tau^*\approx 55$~fs, respectively.
\end{references}
\end{document}